\documentstyle[prb,aps,twocolumn]{revtex}
\begin{document}
\draft
\title{Mean-field theory of the spin-Peierls systems:\\
 Application to CuGeO$_3$}
\author{Mohamed Azzouz and Claude Bourbonnais}
\address{Centre de Recherche en Physique du Solide
et D\'epartement de Physique\\ Universit\'e de Sherbrooke, Sherbrooke
Qu\'ebec,
 J1K 2R1 Canada}
\date{\today}
\maketitle
\begin{abstract}
A mean-field theory of the spin-Peierls systems based on the two dimensional
dimerized Heisenberg model is proposed by introducing an
alternating bond order parameter. Improvements with respect to
previous mean-field results are found in the one-domensional limit for the
ground state and the gap energies.
In two-dimensions, the analysis of the competition between the
antiferromagnetic long range order and the spin-Peierls ordering is
given as a function of the coupling constants.
We show that the lowest
energy gap to be observed does not have a singlet-triplet character
in agreement with the low temperature thermodynamic properties of CuGeO$_3$.

\end{abstract}

The recent discovery of a spin-Peierls (SP) transition in the inorganic
compound  CuGeO$_3$ prompts renewed interest for this  kind of structural
instability. \cite{hase}$^,$\cite{hase1} Evidence for such a
non-magnetic transition
has been exemplified in several ways. The lattice  distorsion  has been
well established by X-ray and elastic neutron
experiments.\cite{ain}$^-$\cite{pouget} The magnetic
susceptibility decreases exponentially showing a gap in the spin
excitations.\cite{hase}$^,$\cite{brill} This is also confirmed by heat
capacity measurements which present a thermally activated component  below
the SP critical temperature.
\cite{oseroff} As shown by  Oseroff {\it et al.},
\cite{oseroff} another interesting feature is the close proximity between
the SP and the antiferromagnetic (AF) ground state for doped CuGeO$_3$
samples.

For the theoretical description of the SP ordered state, the alternating
Heisenberg model has been extensively studied both
numerically\cite{duffy}$^,$\cite{bonner} and analytically.\cite{bula} However
in the latters,
mean-field like decoupling of the Heisenberg interaction in the
quasi-fermion
representation were essentially restricted to the single chain problem.
In the
present work, we introduce a new decoupling for the alternating bond
order parameter which not only improves 1D mean-field results for the
ground state energy and the excitation gap but also allows to treat
the 2D situation,\cite{imada} namely the effect of interchain Heisenberg
exchange interaction. This turns out to be an important ingredient for the
interplay between the AF and the SP states.

We start the analysis with the 2D alternating Heisenberg
model
\begin{eqnarray}
H= &&J\sum_{i,j} {\bf S}_{2i,j}\cdot{\bf S}_{2i+1,j}
  +J'\sum_{i,j} {\bf S}_{2i+1,j}\cdot{\bf S}_{2i+2,j}\cr
&&J_\perp\sum_{i,j} {\bf S}_{i,2j}\cdot{\bf S}_{i,2j+1}
  +J'_\perp\sum_{i,j} {\bf S}_{i,2j+1}\cdot{\bf S}_{i,2j+1}
\label{hamiltonian}
\end{eqnarray}
where $J>0\ (J_\perp>0)$ and $J'>0\ (J'_\perp>0)$ are the intrachain
(interchain)
AF exchange couplings. We then  follow the mean-field approach given in
refs. [12] and [13]. In such a treatment, the spin Hamiltonian
is transformed by means of the generalized Jordan-Wigner (JW) transformation
\cite{azzouz1}
\begin{equation}
\begin{array}{ll}
&S^-_{i,j}= c_{i,j}e^{i\pi(\sum_{\ell= 0}^{i-1}
\sum_{f= 0}^{\infty}n_{\ell,f} + \sum_{f= 0}^{j-1}n_{i,f})}\\
&S^z_{i,j}= c^{\dag}_{i,j}c_{i,j}-{1\over2},
\label{jw}
\end{array}
\end{equation}
where $n_{i,j}=c^{\dag}_{i,j}c_{i,j}$.
The Hamiltonian is then written in Fourier space by taking into account
the bipartite character of the lattice, namely
\begin{eqnarray}
H= {1\over2}\sum_{\bf k}\bigl\{&&Mc^{A\dag}_{\bf k}c^A_{\bf k}
-Mc^{B\dag}_{\bf k}c^B_{\bf k}\cr
&&+e({\bf k})c^{A\dag}_{\bf k}c^B_{\bf k}+
e^*({\bf k})c^{B\dag}_{\bf k}c^A_{\bf k}\big\}.
\label{kham}
\end{eqnarray}
Here $A$ and $B$ label  the two sublattices and $e({\bf k})= J_1{\rm
e}^{ik_x}- J_2{\rm e}^{-ik_x}+J_{\perp1}{\rm e}^{ik_y}+
J_{\perp2}{\rm e}^{-ik_y}$, $M= m(J+J'+J_\perp+J'_\perp)$ with $m$, $Q$,
$Q'$, $P$, $P'$, $J_1$ and $J_2$ are to be defined shortly.
The Hamiltonian (\ref{kham}) is diagonalized using the
following canonical transformation
\begin{equation}
\begin{array}{ll}
&c^A_{\bf k}= u_kf_{\bf k}+v_kd_{\bf k} \\
&c^B_{\bf k}= v^*_kf_{\bf k}-u^*_kd_{\bf k}
\label{transf}
\end{array}
\end{equation}
where $u_k=  e^{i\alpha_k/2}\cos\beta_k$ and
$v_k=  e^{i\alpha_k/2}\sin\beta_k$. $\alpha_k$ and
$\beta_k$ are given by
$$\tan \alpha_k
= \ {(J_1-J_2)\sin k_x+(J_{\perp1}-J_{\perp2})\sin k_y\over
(J_1+J_2)\cos k_x+(J_{\perp1}+J_{\perp2})\cos k_y}
$$
and
\begin{eqnarray}
\tan(2\beta_k)&&= \pm(2M)^{-1} \times \cr
&&\bigl(\
[(J_1+J_2)\cos k_x+(J_{\perp1}+J_{\perp2})\cos k_y]^2 + \cr
&&[(J_1-J_2)\sin k_x+(J_{\perp1}-J_{\perp2})\sin k_y]^2\ \bigr)^{1\over 2}
\nonumber
\end{eqnarray}
where $J_1= J(1+2Q)$, $J_2= J'(1+2Q')$, $J_{\perp1}= J_\perp(1+2P)$ and
$J_{\perp2}= J'_\perp(1+2P')$.
As for the dispersion relation, it is given by
\begin{equation}
E_{\pm}({\bf k})= \pm{1\over2}\bigl\{M^2+|e({\bf k)|^2}\bigr\}^{1/2}
\label{dispersion}
\end{equation}
where ($\pm$) refers to  upper and lower band.
We have introduced the order parameters  $m= 2\langle
S^z\rangle$,
$Q= |\langle c_{2i,j}c_{2i+1,j}^{\dag}\rangle|$,
$Q'= |\langle c_{2i+1,j}c_{2i+2,j}^{\dag}\rangle|$,
$P= |\langle c_{i,2j}c_{i,2j+1}^{\dag}\rangle|$, and
$P'= |\langle c_{i,2j+1}c_{i,2j+2}^{\dag}\rangle|$ for the staggered
magnetization, intra- and interchain bond amplitudes respectively.

Their equilibrium values are obtained from the
minimization of the total free energy leading to
 a set of mean field equations which can be solved exactly.

The ground state wave function corresponds to the case where the lower
band is filled:
\begin{equation}
|\Phi_{GS}\rangle= \prod_{\bf k}d_{\bf k}^{\dag}|0\rangle.
\label{grstate}
\end{equation}
It is formed by the pairs of fermions
$(c_{\bf k}^{A{\dag}},c_{\bf k}^{B{\dag}})$ with the weights
$v_k$ and $-u_k^*$ respectively which
correspond from (\ref{jw}) to pairs of spins denoted
$(\uparrow,\downarrow)_{\bf k}$ in  reciprocal space.
The ground state is a singlet since
$\langle\Phi_{GS}|S^z|\Phi_{GS}\rangle =
\langle\Phi_{GS}|({\bf S}_{tot})^2|\Phi_{GS}\rangle= 0$,
while excited states with a wavevector ${\bf k}'$, relevant to magnetic
susceptibility and specific heat etc, consist in
creating {\it particle-hole} excitations, namely
\begin{equation}
|\Phi_{EX}\rangle= f_{{\bf k}'}^{\dag}d_{{\bf k}'}|\Phi_{GS}\rangle=
\prod_{\bf k}f_{{\bf k}'}^{\dag}d_{{\bf k}'}d_{\bf k}^{\dag}|0\rangle
\label{exstate}
\end{equation}
where the excitation operator can be written  as
\begin{eqnarray}
f_{\bf k}^{\dag}d_{\bf k}\approx u_kv_k^*(S_A^z({\bf k})-S_B^z({\bf k}))
&&-(u_k)^2S_A^+({\bf k})S_B^-({\bf k}) \cr
&&+(v_k^*)^2S_A^-({\bf k})S_B^+({\bf k}),
\label{exoperator}
\end{eqnarray}
where the phase factors in (\ref{jw}) have  been
ignored in order to illustrate that such excitation operator conserves
$S_{tot}^z$ so that
$|\Phi_{EX}\rangle$ is not a triplet. The dispersion relation
for each member of the particle-hole pair is given by
$
{1\over2}\langle\Phi_{EX}|H|\Phi_{EX}\rangle
-{1\over2}\langle\Phi_{GS}|H|\Phi_{GS}\rangle=
E_+({\bf k})$ with an energy gap
\begin{eqnarray}
E_g= {1\over2}&&\bigl\{
m^2(J+J'+J_\perp+J'_\perp)^2\cr
&&+[J_1-J_2]^2+[J_{\perp1}-J_{\perp2}]^2
\bigr\}^{1/2},
\label{gap1}
\end{eqnarray}
at ${\bf k}_0= (0,\pi/2)$.
In the case of uniform exchange interaction, the gap is due to the finite
sublattice magnetization $m$, as discussed previously.
\cite{azzouz1} At zero $m$ and
$J_\perp= J'_\perp$, the gap reduces to
$E_g^{SP}= {1\over2}|J_1-J_2|$ which is due to
the dimerization along the x-axis. The ground state formed by dimers
is the so called SP state. An elementary
excitation will correspond to the destruction of a singlet
$(\uparrow,\downarrow)_{{\bf k}_0}$ from the application of
$f_{{\bf k}_0}^{\dag}d_{{\bf k}_0}$.
Actually, in one dimension one can visualize
$E_g^{SP}$  for the creation of a particle-hole excitation as the energy
required to form a kink in the dimerized chain.
Finally, for longitudinal and transverse alternating exchange interactions,
the gap adds  in quadrature and we find for the particle-hole excitation
\begin{eqnarray}
E_g^{SP}= {1\over2}\bigl\{[J_1-J_2]^2
+[J_{\perp1}-J_{\perp2}]^2
\bigr\}^{1/2}.
\label{gap3}
\end{eqnarray}
Static and uniform spin
susceptibility $\chi$ is equivalent to the compressibility in the  JW
representation. Its evaluation is straightforward and the result is
given by
\begin{eqnarray}
\chi(T)=  && -{1\over2}g^2\mu_B^2 \sum_{\bf k} {\partial n[E_+({\bf k})]\over
\partial
E_+({\bf k})}\cr
    \sim && {1\over2}g^2\mu_B^2{\cal D}(E^{SP}_g)\beta
e^{-\beta E^{SP}_g} \ \ \ \ \ \ (\beta E^{SP}_g\gg1)
\label{chi}
\end{eqnarray}
where $n[x]=  (e^{\beta x} +1)^{-1}$, $g$ is the Land\'e factor
and ${\cal D}(E^{SP}_g)$ is the density of states at the energy gap.
Therefore the lowest
energy gap to be observed in experiments (e.g. in CuGeO$_3$) like magnetic
susceptibility\cite{hase} and specific heat\cite{oseroff} of the condensed SP
state is characterized by the above particle-hole (singlet)
character. Furthermore, in contrast to the critical temperature,
the amplitude of the
zero temperature gap is {\it not} predicted to change when a
magnetic field is applied in agreement with specific heat\cite{oseroff} and
acoustic measurements\cite{mario} performed under low field.

When both possibilities of SP and AF long range order are considered (Eq.\
(\ref{gap1})), the numerical solution of  the  $T=0$ mean field
equations leads to the  phase diagram of
Fig.\ \ref{magnetization} for $(J'/J,J_\perp/J)$ at a fixed $J'_\perp$ value.
Whenever $m\ne 0$, the rotational invariance is broken and the system is in
the AF state while the line boundary is determined when
$m$ vanishes. As for the SP phase, it is defined by
$m=0$ and an energy gap solely due to dimerization. In the insert of Fig.\
\ref{magnetization}, the magnetization is  displayed as a function of $J'/J$
for
$J_\perp=J'_\perp=.1J$. Therefore whenever the dimerization becomes
sufficiently small, the SP ordered state is no longer stable and a magnetic
ordering is favored; an increase of the interchain exchanges also
favors the magnetic ordering. This is consistent with the situation found in
real systems like quasi-1D organic materials
\cite{PF6} where the application of hydrostatic pressure is well known to
promote such a crossover.

Focussing now on the ground state energy
$E_{GS}=\langle\Phi_{GS}|H|\Phi_{GS}\rangle$ in the $m=0$ SP phase, one
can
compare in Figure \ref{fig1d} the present mean-field approach with the previous
results obtained by Bulaesvskii\cite{bula} in the 1D limit. Thus the choice
of an alternated order parameter ($Q\ne Q'$) gives rise to a better
estimation of the ground state energy. In the uniform Heisenberg limit of
the model, $J=J'$, the  dispersion
relation $E_+(k)=(1+2Q)|\sin k|$ becomes gapless with
$1 +2Q \approx 1.63$
in fair agreement with the  exact result.\cite{cloiseaux}

As far as the gap is concerned in
this 1D limit (insert of the Figure \ref{fig1d}), the present approach
leads to
\begin{equation}
E_g^{SP}= E_0 + C\mid 1-J'/J\mid^\alpha,
\label{gap1d}
\end{equation}
with $C\simeq .8J$ and the exponent $\alpha \simeq .71$ which is close to the
Cross and Fisher \cite{cross} value ($\alpha=2/3$),  RG
results ($\alpha= .76$) \cite{fields} and exact diagonalization
($\alpha= .79\pm .06$).\cite{kiyomi} The present calculation however
predicts a
finite jump $E_0\simeq .19J$ for the energy gap once $\mid J-J'\mid$ is non
zero; a result not yet confirmed by an exact numerical calculation
probably  due to finite size effects as $\mid J-J'\mid\to 0 $.\cite{duffy}

When the effect of interchain exchange coupling is included (Figure
\ref{figgap2}), one can extract in the region where the SP phase is stable
(for $J_\perp/J<.45$, the AF phase is dominant for all $|J-J'|$)
the universal value $\alpha\simeq .66$ for the exponent, while the constant
$E_0$ decreases monotonously with $J_\perp$ and $C\simeq.81J$.
As a 2D mean-field result, this value of $\alpha$ is closer to the one
obtained by Cross and Fisher.\cite{cross}

In summary, we have proposed a mean-field theory of the 2D dimerized
Heisenberg model with magnetic and alternating bond order parameters
for the description of the ordered spin-Peierls
state and its competition with the antiferromagnetic order.
In the 1D limit, the ground state energy and the particle-hole excitation
gap
profiles with dimerization show marked improvements with respect to previous
mean-field results. As far as the excitation gap is concerned, particle-hole
excitations should dominate the thermodynamics at low temperature and this,
consistently with recent  measurements on specific heat and magnetic
susceptibillity for the CuGe0$_3$ material.

Enlightning discussions with M. Plumer and M. Ain were very
helpful. This work was supported by the Natural Sciences and
Engineering Research Council of Canada (NSERC) and the Fonds pour
la formation de chercheurs et l'aide a la recherche from the Government of
Qu\'ebec (FCAR).
%

%
%%%%%%%%%%%%%%%%%%%%%%%%%%%%%%%%%%%%%%%%%%%%%%%%%%%%%%%%%%%%%%%%%%%%%%%%%%%%
\begin{figure}
\caption{The phase digram $(J'/J,J_\perp/J)$ is drawn for
$J'_\perp= .08J$. In the insert,
the magnetization $m$ is reported as a function of $J'/J$ for
$J'_\perp= J_\perp= .1J$.}
\label{magnetization}
\end{figure}
%%%%%%%%%%%%%%%%%%%%%%%%%%%%%%%%%%%%%%%%%%%%%%%%%%%%%%%%%%%%%%%%%%%%%%%%%%%%
%
%
%%%%%%%%%%%%%%%%%%%%%%%%%%%%%%%%%%%%%%%%%%%%%%%%%%%%%%%%%%%%%%%%%%%%%%%%%%%%
\begin{figure}
\caption{The ground state energy as a function of $|J'-J|$ in the present
mean-field approximation compared to the Bulaevskii's results (B). In
the insert, the same comparison for 1D SP gap.}
\label{fig1d}
\end{figure}
%%%%%%%%%%%%%%%%%%%%%%%%%%%%%%%%%%%%%%%%%%%%%%%%%%%%%%%%%%%%%%%%%%%%%%%%%%%%
%
%
%%%%%%%%%%%%%%%%%%%%%%%%%%%%%%%%%%%%%%%%%%%%%%%%%%%%%%%%%%%%%%%%%%%%%%%%%%%%
\begin{figure}
\caption{The energy gap $E_g^{SP}$ as a function of
$\mid J-J'\mid $ for $J_\perp=.15J$ (curve 1) and  $J_\perp=  0.1J$
(curve 2). The curve 3 gives the energy gap which would be obtained
in the absence of antiferromagnetism ($m=0$) for $J_\perp= .15J$.
In the insert, the magnetization is
reported for $J_\perp= .15J$ (curve 1) and  $J_\perp= .1J$
(curve 2).}
\label{figgap2}
\end{figure}
%%%%%%%%%%%%%%%%%%%%%%%%%%%%%%%%%%%%%%%%%%%%%%%%%%%%%%%%%%%%%%%%%%%%%%%%%%%%
%
%
\end{document}